\documentclass[11pt]{article}

\usepackage[a4paper,margin=1in]{geometry}
\usepackage{times}
\usepackage{booktabs}
\usepackage{array}
\usepackage{longtable}
\usepackage{pdflscape}
\usepackage{graphicx}
\usepackage{hyperref}
\usepackage{amsmath}
\usepackage{enumitem}
\usepackage{caption}
\usepackage{titlesec}
\usepackage{authblk}
\usepackage[numbers]{natbib}
\usepackage{tabularx}
\usepackage{xurl}

\hypersetup{
  colorlinks=true,
  linkcolor=black,
  citecolor=black,
  urlcolor=black
}

\title{\textbf{Benchmark-Based Comparative Assessment of Publicly Benchmarked Indian Foundation Models:\\ A Capability and Evaluation-Maturity Framework}}

\usepackage{authblk}

\author[1]{Avinash Agarwal\thanks{Corresponding author. Deputy Director General. Email: avinash.70@gov.in}}
\author[1]{Vridhi Jain\thanks{Intern}}

\affil[1]{Unique Identification Authority of India, New Delhi, India}
\date{August 2026}

\begin{document}

\maketitle

\begin{abstract}
\textbf{Purpose}: Governments increasingly fund indigenous foundation models to strengthen national AI capability, digital sovereignty, and multilingual computing. This paper assesses India's foundation-model ecosystem and examines whether apparent capability gaps in public benchmark evidence may also reflect gaps in evaluation maturity.

\textbf{Design/methodology/approach}: The paper presents a structured, benchmark-based comparative assessment of Indian foundation models against global frontier and comparable-scale models across eight capability domains: general-purpose reasoning, coding and software engineering, agentic AI and computer use, cybersecurity, vision and image understanding, video and multimodal understanding, scientific research, and Indic language capability. Using only publicly reported results, it proposes an exploratory four-dimension Benchmark Maturity Index (BMI), scoring each domain on standardization, participation, independent verification, and national coverage.

\textbf{Findings}: Indian models achieve strong scores on established benchmarks such as MMLU and MATH-500. However, these are now widely regarded as saturated, and frontier developers no longer report them. Indian models participate far less frequently in newer, agentic, and domain-specialized evaluations, and participation is highly uneven across organizations. Sarvam AI reports the broadest coverage by a substantial margin. The BMI refines, and in some cases revises, the maturity judgments a purely descriptive review would produce.

\textbf{Practical implications}: Many apparent capability gaps cannot be distinguished, on available evidence, from evaluation-ecosystem gaps, with direct implications for how national AI programs should design monitoring and funding criteria.

\textbf{Originality}: The paper proposes the BMI as a reusable instrument for scoring evaluation-ecosystem maturity at the domain level and demonstrates its application to the Indian foundation-model ecosystem.
\\

\noindent\textbf{Keywords:} generative AI, Indian foundation models, benchmarking, comparative evaluation, benchmark maturity, Benchmark Maturity Index, AI evaluation, sovereign AI
\end{abstract}

\section{Introduction}
\label{sec:intro}

Generative artificial intelligence has moved rapidly from text generation toward a broad set of capabilities. These now include reasoning, software engineering, multimodal understanding, image and video generation, scientific discovery, and autonomous agentic behavior \citep{hendrycks2021mmlu, chen2021humaneval}. As these capabilities have matured, national governments have increasingly treated frontier AI as a strategic technology. They fund indigenous foundation models to build technological capability, digital sovereignty, multilingual computing capacity, and domain-specific innovation. India's IndiaAI Mission is one such program. It supports the development of sovereign foundation models trained on Indian data and aligned with India's linguistic and social context. India has adopted a sector-led, relatively light-touch approach to AI governance, which promotes innovation but risks policy fragmentation and incomparable evaluation practices across domains \citep{agarwal2026federated}.As of February 2026, the IndiaAI Innovation Centre (Foundation Models) pillar has selected 12 organizations and consortia for support, including both private companies and academic consortia \citep{pib2026indiaai}.

Assessing the progress of a national AI ecosystem is not straightforward. It requires more than identifying which models exist. It requires understanding which capability domains are represented, how models compare with global frontier systems, and where genuine capability gaps remain as opposed to gaps in public disclosure. This distinction is the central concern of this paper. It matters because different organizations evaluate their models using different datasets, benchmark suites, evaluation harnesses, and reporting practices. Some organizations publish extensive benchmark tables at every release. Others disclose only selected results, or none at all. A national capability assessment that does not account for this will be incomplete at best. At worst, it will mistake an evaluation-reporting gap for a capability gap.

This paper undertakes a structured, benchmark-based comparative assessment of publicly benchmarked Indian foundation models with this distinction as its organizing concern. Rather than focusing on any single model family, we compare representative Indian models against global frontier models and against global models of a comparable scale, across eight capability domains. We restrict our analysis to publicly reported benchmark results drawn from technical reports, model cards, and benchmark leaderboards. We distinguish between developer-reported scores and independently verified scores throughout. We then ask a second, deliberately separate question: not only how capable is each domain's best Indian model, but how mature and independently verifiable is the evaluation ecosystem within which that capability was measured.

\subsection{Contributions}

This paper makes four contributions.

\begin{enumerate}[leftmargin=*]
\item To our knowledge, this is among the first structured, eight-domain, benchmark-based comparative assessments of publicly benchmarked Indian foundation models against global frontier and comparable-scale models.
\item We propose a graded, three-tier working definition of an ``Indian-developed AI model.'' This distinguishes fully indigenous development from India-led development with global components, and from India-adapted foreign base models. We apply this definition to the 12 organizations selected under the IndiaAI Innovation Centre.
\item We propose an exploratory four-dimension Benchmark Maturity Index (BMI). It scores standardization, participation, independent verification, and national coverage at the level of a capability domain. We show that it refines, and in two cases revises, the maturity judgments that a purely qualitative review would produce. We present this as a framework for further development rather than a validated instrument.
\item We identify and analyze eight cross-cutting properties of the current benchmark ecosystem that materially affect any national capability comparison. These include benchmark saturation, the absence of a common benchmark set across providers, and the divergence between the benchmarks that Indian and global developers choose to report.
\end{enumerate}

The remainder of this paper is organized as follows. Section~\ref{sec:related} situates this work relative to existing literature. Section~\ref{sec:method} describes our methodology, including model and benchmark selection and our working definition of an Indian-developed model. Section~\ref{sec:indiaai} maps the IndiaAI-supported ecosystem. Section~\ref{sec:domains} presents the comparative assessment across eight capability domains. Section~\ref{sec:bmi} introduces the Benchmark Maturity Index. Section~\ref{sec:discussion} discusses cross-cutting patterns, limitations, and policy implications. Section~\ref{sec:conclusion} concludes.

\section{Related Work}
\label{sec:related}

Systematic benchmarking of foundation models has become a distinct area of work. The Holistic Evaluation of Language Models (HELM) project \citep{liang2023helm} was among the first efforts to evaluate a broad set of language models across a standardized battery of scenarios and metrics. It was explicitly motivated by the observation that model developers report inconsistent and self-selected subsets of benchmark results. Chatbot Arena \citep{chiang2024chatbotarena} took a complementary approach, using pairwise human preference voting to rank models independently of developer-reported scores. Independent commercial platforms, including Artificial Analysis \citep{artificialanalysis2026} and Epoch AI \citep{epochai2026}, have since extended this practice. They maintain continuously updated leaderboards that track frontier model releases across cost, speed, and capability dimensions. Model Evaluation and Threat Research (METR) \citep{metr2026} has focused specifically on independently verifying agentic and autonomous capability claims, an area where self-reported benchmark scores are particularly difficult to compare.

This body of work establishes that benchmark fragmentation and self-reporting bias are general problems in the foundation model ecosystem. They are not problems specific to any single country's models. However, existing benchmarking efforts are largely organized around individual models or a global leaderboard, rather than around the question of how a national AI ecosystem compares with the global frontier. This is the gap the present paper addresses. Separately, composite indices have been proposed for assessing national AI regulatory readiness \citep{agarwal2026arri}, addressing the legal and institutional preparedness to govern AI. The present study addresses a complementary but distinct question: the maturity of the evaluation ecosystem itself.We are not aware of a prior structured, multi-domain benchmark comparison of Indian foundation models specifically, though individual Indian model releases, such as Sarvam AI's technical reports, do report comparative benchmark tables against selected global models \citep{sarvam2026}.

Separately, a growing literature documents specific problems with benchmark reliability itself. These include data contamination \citep{jimenez2024swebench, shi2024contamination}, benchmark saturation as top models approach ceiling performance \citep{kiela2021dynabench}, and the sensitivity of reported scores to prompting and harness configuration. We draw on this literature in Section~\ref{sec:discussion}. We argue that these problems are not incidental to a national capability assessment but are central to interpreting it correctly.

\section{Methodology}
\label{sec:method}

\subsection{Model and Benchmark Selection}

We compare models across three groups.

The first group consists of \textbf{global frontier models}: GPT-5.6 \citep{openai2026gpt56}, Claude Opus 5 \citep{anthropic2026opus5}, Kimi K3 \citep{moonshot2026kimik3}, and Qwen3.8-Max \citep{alibaba2026qwen38max}. These were selected as the four highest-ranked models on third-party frontier leaderboards (Chatbot Arena and Artificial Analysis) as of August 2026 that also had publicly available technical reports. The BMI's Participation dimension is scored against this four-model sample and is therefore sensitive to sample composition.

The second group consists of \textbf{global active-parameter comparable models}. We define this group by active parameter count during inference rather than raw total parameter count, because mixture-of-experts (MoE) architectures make total parameter counts misleading as a proxy for model scale. Specifically, we include models whose active parameter count falls in the approximate range of 12--50 billion parameters. The models in this group are: Inkling \citep{tml2026inkling} (MoE; approximately 975B total parameters, approximately 41B active), Qwen3.6-27B \citep{alibaba2026qwen36} (dense; 27B parameters), and Nemotron 3 Super \citep{nvidia2026nemotron3} (MoE; approximately 120B total parameters, approximately 12B active). This group is heterogeneous: active parameter count is an imperfect proxy for inference-time compute, which also depends on architecture, attention mechanism, sequence length, and implementation. Comparisons within this group should be interpreted with caution.

The third group consists of \textbf{representative Indian foundation models}, identified from public technical reports and organizational announcements: Sarvam-105B (MoE; 106B total parameters, 10.3B active parameters) and Sarvam-30B (MoE; 32B total parameters, 2.4B active parameters) \citep{sarvam2026}, Param2 (MoE; approximately 17B total parameters, approximately 2.4B active) \citep{bharatgen2025param2}, and Krutrim-2 \citep{krutrim2026}. We note that Param2's active parameter count of approximately 2.4B places it well below the 12--50B active-parameter range used to define the comparable-scale group. It is included in the Indian group as an IndiaAI-supported model, not because it is parameter-comparable to the global comparable-scale group. We also note domain-specific Indian systems in scientific research (Shodh AI's Project Skanda, IntelliHealth's NeuroDX, ZenteiQ's BrahmAI) and video generation (Avataar AI's Varya). Section~\ref{sec:indiaai} provides a broader mapping of the IndiaAI-supported ecosystem.

Representative benchmarks were identified separately for each of eight capability domains: general-purpose reasoning and knowledge, coding and software engineering, agentic AI and computer use, cybersecurity, vision and image understanding, video and multimodal understanding, scientific research, and Indic language capability. Where an originally identified benchmark did not have publicly reported results for the model set under review, a substitute was selected. This is documented in Section~\ref{sec:video}.

Benchmark scores were extracted from publicly available sources: model technical reports, system cards, official benchmark leaderboards, and comparative benchmark tables reported by model developers. We distinguish throughout between \emph{developer-reported} scores (published by the model developer) and \emph{independently verified} scores (published by a third-party evaluator or independent leaderboard). We did not run any evaluation ourselves. Every score in this paper is therefore a reported figure, not an independently reproduced measurement. This is discussed in Section~\ref{sec:limitations}.

Data collection was conducted in August 2026. Foundation model benchmark scores change frequently. Readers should treat the figures in this paper as a snapshot and verify current figures against cited primary sources. 

\subsection{Defining an Indian-Developed AI Model}

Not all models built by Indian organizations are equally indigenous. This distinction matters for both fair comparison and funding policy. We propose a three-tier working definition.

\textbf{Tier 1 (fully indigenous):} Models trained from scratch by an Indian organization, using India-based or India-controlled compute infrastructure, on datasets that include a significant share of Indian-sourced data. Terms such as ``India-controlled compute'' and ``significant share'' are qualitative; operationalizing them precisely would require training details that are not always publicly available. Sarvam-105B and Sarvam-30B meet this definition: both were trained from scratch on internally curated datasets using IndiaAI Mission compute at Yotta's Shakti H100 cluster in India \citep{sarvam2026}. BharatGen/Param2 is provisionally classified as Tier 1, given its association with AIRAWAT national compute \citep{bharatgen2025param2}, pending independent verification of all criteria.

\textbf{Tier 2 (India-led, global components):} Models developed by an Indian organization that may use foreign cloud compute, foreign training frameworks, or a mixture of global and Indian datasets.

\textbf{Tier 3 (India-adapted):} Foreign base models that are fine-tuned or otherwise adapted by an Indian organization for Indian languages or use cases, without ground-up development in India. Avataar AI's Varya is a Tier 3 model: it is distilled from Alibaba's open-source Wan 2.2 and adapted for Indian cultural contexts, rather than being trained from scratch \citep{mehta2026avataar}.

This tiering serves two purposes. First, it provides a defensible basis for comparison. A Tier 1 model faces materially different constraints in compute, data availability, and engineering capacity than a Tier 3 model built on an existing foreign foundation. Second, it can inform differentiated funding criteria. Tier 1 development plausibly requires more sustained public investment in compute and data infrastructure, while Tier 3 work may require less. This framing extends existing policy language: the IndiaAI Mission's 2025 Call for Proposals explicitly favors models trained on Indian datasets and aligned with India's linguistic and societal context.

\subsection{Scope and Exclusions}

This assessment does not include every Indian AI initiative or every global model release. Models were included on the basis of public availability of technical documentation as of the data collection date. Proprietary models with no publicly disclosed benchmark results were excluded from the comparative tables. Their existence is noted qualitatively in Section~\ref{sec:concentration} and the full IndiaAI-supported population is catalogued in Section~\ref{sec:indiaai}. We do not evaluate model safety, alignment, deployment cost, or inference efficiency. This paper is concerned exclusively with publicly reported task-capability benchmark performance.

Because our conclusions are conditioned on public disclosure, they apply to the publicly benchmarked subset of the Indian ecosystem. They do not necessarily characterize the Indian AI ecosystem as a whole. Models that exist but do not publish benchmark results are invisible to this methodology.

\section{The IndiaAI-Supported Foundation Model Ecosystem}
\label{sec:indiaai}

The Government of India's IndiaAI Innovation Centre (Foundation Models) pillar has selected 12 organizations and consortia for the development of large and small language models based on Indian datasets. The February 2026 Press Information Bureau (PIB) disclosure provides details of the compute and non-compute support extended to each \citep{pib2026indiaai}. Table~\ref{tab:indiaai} maps these 12 organizations, their known models, and their publicly available benchmark evidence as of August 2026.

\begin{table}[h]
\centering
\small
\caption{IndiaAI Innovation Centre (Foundation Models): 12 Supported Organizations and Public Benchmark Evidence}
\label{tab:indiaai}
\begin{tabular}{llcc}
\toprule
\textbf{Organization} & \textbf{Known Model(s)} & \textbf{Tier} & \textbf{Public Benchmarks?} \\
\midrule
Sarvam AI & Sarvam-30B, Sarvam-105B & 1\textsuperscript{\ddag} & Yes (broad) \\
IIT Bombay / BharatGen & Param2 & 1\textsuperscript{\dag} & Limited \\
Avataar AI & Varya (video generation) & 3\textsuperscript{\S} & Limited \\
Shodh AI & Project Skanda & --- & None found \\
IntelliHealth & NeuroDX & --- & None found \\
ZenteiQ & BrahmAI & --- & None found \\
Soket AI & --- & --- & None found \\
Gnani AI & --- & --- & None in surveyed domains\textsuperscript{\P} \\
Gan AI & --- & --- & None found \\
GenLoop & --- & --- & None found \\
Fractal Analytics & --- & --- & None found \\
Tech Mahindra Maker's Lab & --- & --- & None found \\
\bottomrule
\end{tabular}
\\[2pt]
\footnotesize Tier assignments follow the definition in Section~\ref{sec:method}. A dash indicates insufficient public documentation to assign a tier. ``None found'' means no standardized benchmark results were located in public technical reports as of August 2026. \textsuperscript{\dag}Provisionally classified; see note in text. \textsuperscript{\ddag}Sarvam-105B and Sarvam-30B were trained from scratch, using IndiaAI Mission compute (Yotta's Shakti H100 cluster), on internally curated datasets; they meet the stated Tier 1 criteria. \textsuperscript{\S}Varya is distilled from Alibaba's open-source Wan 2.2 and adapted for Indian cultural contexts; it meets the Tier 3 definition. \textsuperscript{\P}Gnani AI reports a result on the Berkeley Function-Calling Leaderboard (BFCL) v3 (37.99\%), a function-calling benchmark outside the eight capability domains assessed in this study.
\end{table}

We note two distinctions. First, the Tier 1 classification for BharatGen/Param2 is provisional. BharatGen is associated with AIRAWAT national compute infrastructure, which satisfies one element of the Tier 1 definition. However, we have not independently verified that all Tier 1 requirements (training from scratch, India-controlled compute, significant Indian-sourced data) are met. The relevant primary documentation should be consulted for confirmation.

Second, in addition to these 12 IndiaAI-supported organizations, our benchmark comparison includes \textbf{Krutrim AI} (Krutrim-2), which is an Indian foundation-model developer not included in the official 12-organization IndiaAI Innovation Centre list. Krutrim is included in the comparative assessment because it has publicly released a model and reported benchmark results. Table~\ref{tab:other_indian} records this.

\begin{table}[h]
\centering
\small
\caption{Other Indian Foundation Models Included in Comparative Assessment}
\label{tab:other_indian}
\begin{tabularx}{\textwidth}{l l c c X}
\toprule
\textbf{Organization} & \textbf{Model} & \textbf{Tier} & \textbf{IndiaAI-Supported?} & \textbf{Public Benchmarks?} \\
\midrule
Krutrim AI & Krutrim-2 & 2 & No & Limited (3 domains: General Purpose AI, Coding, Indic Language) \\
\bottomrule
\end{tabularx}
\end{table}

Of the 12 IndiaAI-supported organizations, only Sarvam AI publishes benchmark results across multiple capability domains. BharatGen/Param2 reports limited results. The remaining 10 either have not released models publicly or have not disclosed standardized benchmark scores within the domains assessed in this study (though Gnani AI reports a result on Berkeley BFCL v3, outside our domain scope). Krutrim, though not IndiaAI-supported, reports results across three domains: MMLU (general purpose), HumanEval (coding), and BharatBench (Indic language). This observation is central to Section~\ref{sec:concentration}. It also motivates our argument that benchmark disclosure practices should be part of any evaluation of publicly funded AI programs.

We emphasize that the absence of public benchmark results does not establish the absence of model capability. Organizations may have working models that have not been publicly benchmarked, or may have results that have not been disclosed. The table records public evidence only.

\section{Comparative Assessment Across Capability Domains}
\label{sec:domains}

Throughout this section, we use the phrase ``not reported'' to mean that no publicly available benchmark score was found for the model tier in question. This may reflect any of several states: the evaluation was not run, was run but not disclosed, was excluded for strategic reasons, or was not applicable. We do not distinguish between these states because the public record does not allow us to do so.

\subsection{General-Purpose Foundation Models}

General-purpose foundation models are evaluated on knowledge retrieval, reasoning, and mathematical problem solving. We use MMLU \citep{hendrycks2021mmlu}, MMLU-Pro \citep{wang2024mmlupro}, GPQA Diamond \citep{rein2023gpqa}, Humanity's Last Exam (HLE) \citep{hle2025}, MATH-500 \citep{lightman2024math500}, and ARC-AGI-2 and ARC-AGI-3 \citep{chollet2025arcagi2,arcprize2026arcagi3}.

\begin{table}[h]
\centering
\small
\caption{General-Purpose Foundation Models: Best Reported Score by Model Tier}
\label{tab:general}
\begin{tabular}{lccc}
\toprule
\textbf{Benchmark} & \textbf{Best Surveyed Frontier} & \textbf{Best Surveyed Comparable} & \textbf{Best Indian} \\
\midrule
MMLU & Not reported & Not reported & Sarvam-105B, 90.6 \\
MMLU-Pro & Claude Opus 5, 91.6 & Not reported & Sarvam-105B, 81.7 \\
GPQA Diamond & GPT-5.6 Sol, 94.1 & Inkling, 87.2 & Sarvam-105B, 78.7 \\
HLE (with tools) & Claude Opus 5, 64.7 & Inkling, 46.0 & Sarvam-105B, 11.2 \\
MATH-500 & Not reported & Not reported & Sarvam-105B, 98.6 \\
ARC-AGI-3 & Claude Opus 5, 30.2 & Not reported & Not reported \\
ARC-AGI-2 & GPT-5.6 Sol, 92.5 & Not reported & Not reported \\
\bottomrule
\end{tabular}
\end{table}

Indian models achieve strong scores on MMLU and MATH-500, two widely used benchmarks for general-purpose knowledge and mathematical reasoning. However, these benchmarks have become less discriminative among stronger models as performance has approached the upper end of these benchmarks \citep{kiela2021dynabench}. Frontier developers in this review do not report scores on them, consistent with the saturation pattern discussed in Section~\ref{sec:saturation}. Strong performance on saturated benchmarks demonstrates competence but does not establish frontier competitiveness. MMLU-Pro \citep{wang2024mmlupro}, a harder successor benchmark, offers a partial exception: Sarvam-105B reports 81.7 and Claude Opus 5 reports 91.6, giving a frontier-Indian comparison point that plain MMLU no longer provides.

On GPQA Diamond, the one benchmark with reported scores across all three tiers, Sarvam-105B (78.7) trails the frontier leader by roughly 15 points. The best comparable-scale score is Inkling at 87.2, which also exceeds Sarvam-105B by a meaningful margin. On HLE with tools, Inkling (46.0) substantially outperforms Sarvam-105B (11.2), indicating a significant gap between Indian models and even comparable-scale global models on harder reasoning benchmarks. No Indian model reports a score on ARC-AGI-2 or ARC-AGI-3. These are newer benchmarks introduced specifically because earlier benchmarks had become uninformative at the frontier.

\subsection{Coding and Software Engineering}
\label{sec:coding}

Coding benchmarks range from foundational code generation (HumanEval \citep{chen2021humaneval}, MBPP \citep{austin2021mbpp}) to real-world, multi-step software engineering tasks (SWE-bench Pro \citep{jimenez2024swebench}, Terminal-Bench 2.1).

\begin{table}[h]
\centering
\small
\caption{Coding and Software Engineering: Best Reported Score by Model Tier}
\label{tab:coding}
\begin{tabular}{lccc}
\toprule
\textbf{Benchmark} & \textbf{Best Surveyed Frontier} & \textbf{Best Surveyed Comparable} & \textbf{Best Indian} \\
\midrule
HumanEval & Not reported & Not reported & Sarvam-30B, 92.1 \\
MBPP & Not reported & Not reported & Sarvam-30B, 92.7 \\
LiveCodeBench & GPT-5.6 Sol, 82.6 & Not reported & Sarvam-105B, 71.7 \\
SWE-bench Pro & Claude Opus 5, 79.2 & Inkling, 54.3 & Not reported \\
DeepSWE & GPT-5.6 Sol, 73.0 & Not reported & Not reported \\
Terminal-Bench 2.1 & GPT-5.6 Sol, 88.8 & Inkling, 63.8 & Not reported \\
Vibe Code Bench v1.1 & GPT-5.6 Sol, 80.5 & Not reported & Not reported \\
\bottomrule
\end{tabular}
\end{table}

This domain shows a sharper divide than general-purpose reasoning. Indian models report strong scores on foundational code-generation benchmarks (HumanEval, MBPP, LiveCodeBench). No Indian model reports a result on any of the four agentic or multi-step software engineering benchmarks in this review. Comparable-scale models also participate only partially, appearing on SWE-bench Pro and Terminal-Bench 2.1 but not on DeepSWE or Vibe Code Bench.

The available public evidence therefore shows Indian benchmark participation concentrated in foundational coding, with no publicly reported results in agentic coding. Whether this reflects a genuine capability gap or only a reporting gap cannot be determined from the available evidence.

Two further observations on the global coding table are warranted. First, HumanEval and MBPP show signs of saturation: multiple models across tiers achieve scores above 90, and frontier developers no longer report them. The frontier evaluation focus has shifted to harder agentic benchmarks such as SWE-bench Pro and Terminal-Bench. Second, Claude Opus 5 leads SWE-bench Pro in this table at 79.2\%, above Qwen3.8-Max (67.7\%) and GPT-5.6 Sol (73.0\%). Despite this, Claude Opus 5 does not appear as the top scorer on the remaining coding benchmarks, because Anthropic does not publish scores on DeepSWE, Terminal-Bench 2.1, or Vibe Code Bench v1.1. The pattern illustrates how selective benchmark reporting can make a capable model appear absent from a domain in which it is competitive. We return to this domain's benchmark fragmentation in Section~\ref{sec:bmi}.

\subsection{Agentic AI and Computer Use}
\label{sec:agentic}

Agentic AI benchmarks evaluate a model's ability to autonomously navigate web interfaces, operate computer environments, and complete multi-step tasks. We use BrowseComp, OSWorld-Verified, and OSWorld 2.0.

\begin{table}[h]
\centering
\small
\caption{Agentic AI and Computer Use: Best Reported Score by Model Tier}
\begin{tabular}{lccc}
\toprule
\textbf{Benchmark} & \textbf{Best Surveyed Frontier} & \textbf{Best Surveyed Comparable} & \textbf{Best Indian} \\
\midrule
BrowseComp & Kimi K3, 91.2 & Inkling, 77.1 & Sarvam-105B, 49.5 \\
OSWorld-Verified & Qwen3.8-Max, 86.1 & Not reported & Not reported \\
OSWorld 2.0 & Claude Opus 5, 70.6 & Not reported & Not reported \\
\bottomrule
\end{tabular}
\end{table}

This domain is notable because BrowseComp is the only benchmark in this study where all three model tiers report results on the same benchmark. This makes it the strongest basis for direct three-tier comparison anywhere in the paper. Sarvam-105B (49.5) and Sarvam-30B (35.5) both report BrowseComp scores, trailing the frontier leader Kimi K3 (91.2) and comparable-scale leader Inkling (77.1) by substantial margins. The gap is large but the existence of Indian participation on a shared benchmark is itself significant. It directly supports the paper's central argument: meaningful comparison is possible wherever standardized evaluation exists and all tiers participate.

On OSWorld-Verified and OSWorld 2.0, only frontier models report scores. No comparable-scale or Indian model participates. This pattern of partial participation mirrors other domains.

\subsection{Cybersecurity}

Cybersecurity evaluation uses task-based benchmarks measuring vulnerability discovery and exploitation capability: CyberGym, ExploitBench, ExploitGym, and CVE-Bench.

\begin{table}[h]
\centering
\small
\caption{Cybersecurity: Best Reported Score by Model Tier}
\begin{tabular}{lccc}
\toprule
\textbf{Benchmark} & \textbf{Best Surveyed Frontier} & \textbf{Best Surveyed Comparable} & \textbf{Best Indian} \\
\midrule
CyberGym & GPT-5.6 Sol, 83.6 & Not reported & Not reported \\
ExploitBench (Cap\%) & GPT-5.6 Sol / Claude Opus 5, $\approx$70 & Not reported & Not reported \\
ExploitGym (2h / 6h) & GPT-5.6 Sol, 25 / 35 & Not reported & Not reported \\
CVE-Bench (pass@1) & GPT-5.6 Sol, 93 & Not reported & Not reported \\
\bottomrule
\end{tabular}
\end{table}

Cybersecurity shows no cross-tier comparability in this review. Reported results are confined to the frontier tier. Even within that tier, participation is incomplete: only GPT-5.6 Sol and Claude Opus 5 report scores. No comparable-scale or Indian model reports any cybersecurity benchmark result. This makes cybersecurity the domain with the least publicly available evidence for any comparative claim. No conclusion about Indian cybersecurity AI capability can be drawn from the available public benchmark record.

\subsection{Vision and Image Understanding}

We use MMMU \citep{yue2024mmmu} and MMMU-Pro. Dedicated image-generation benchmarks (GenEval \citep{ghosh2023geneval}, HEIM \citep{lee2023heim}) were reviewed but excluded from the comparative table. The models under review report multimodal understanding results, not native image-generation results.

\begin{table}[h]
\centering
\small
\caption{Vision and Image Understanding: Best Reported Score by Model Tier}
\begin{tabular}{lccc}
\toprule
\textbf{Benchmark} & \textbf{Best Surveyed Frontier} & \textbf{Best Surveyed Comparable} & \textbf{Best Indian} \\
\midrule
MMMU & GPT-5.6 Sol, 88.8 & Not reported & Not reported \\
MMMU-Pro (with tools) & GPT-5.6 Sol, 84.6 & Inkling, 73.5 & Not reported \\
Design Arena & Not reported & Not reported & Not reported \\
\bottomrule
\end{tabular}
\end{table}

Participation is limited even within the frontier tier. No Indian model reports a result on any vision-understanding benchmark reviewed here, though at least one Indian multimodal system has been publicly announced. This reflects a general scarcity of disclosed vision-understanding results across the ecosystem, not a gap specific to Indian models.

\subsection{Video and Multimodal Understanding}
\label{sec:video}

The two benchmarks originally identified for this domain, VideoMMMU and MVBench, did not have publicly reported results for the surveyed models. We therefore substitute Video-MME \citep{fu2024videomme} and MMVU. We report this substitution explicitly. However, we note that the need for substitution demonstrates limited public reporting coverage for our model set, not necessarily immaturity of the underlying benchmark ecosystem. Mature benchmarks can have low model participation.

\begin{table}[h]
\centering
\small
\caption{Video and Multimodal Understanding: Best Reported Score by Model Tier}
\begin{tabular}{lccc}
\toprule
\textbf{Benchmark} & \textbf{Best Surveyed Frontier} & \textbf{Best Surveyed Comparable} & \textbf{Best Indian} \\
\midrule
Video-MME & Kimi K3, 90.0 & Not reported & Not applicable \\
MMVU & Kimi K3, 82.1 & Not reported & Not applicable \\
\bottomrule
\end{tabular}
\end{table}

The Indian model identified for this area, Varya (Avataar AI), is a dedicated video-\emph{generation} system, not a video-\emph{understanding} model. It is therefore not comparable on the benchmarks above. We mark this ``not applicable'' rather than as a zero, since the mismatch is one of task definition. India has no publicly benchmarked video-understanding model in our survey. No comparable-scale model reports a result on either benchmark.

Video generation and video understanding are fundamentally different tasks; Varya's existence is not evidence of Indian video-understanding capability.

\subsection{Scientific Research}

Scientific research systems are evaluated using domain-specific tasks rather than a single unified benchmark. We reviewed ProtocolQA, LAB-Bench, ProteinGym, and PaperBench.

\begin{table}[h]
\centering
\small
\caption{Scientific Research: Best Reported Score by Model Tier}
\begin{tabular}{lccc}
\toprule
\textbf{Benchmark} & \textbf{Best Surveyed Frontier} & \textbf{Best Surveyed Comparable} & \textbf{Best Indian} \\
\midrule
ProtocolQA (Open-Ended) & GPT-5.6 Sol, 43.5 & Not reported & Not reported \\
PaperBench & Qwen3.8-Max, 93.0 & Not reported & Not reported \\
LAB-Bench & Not reported & Not reported & Not reported \\
ProteinGym & Not reported & Not reported & Not reported \\
\bottomrule
\end{tabular}
\end{table}

Frontier models show limited but non-zero participation in this domain. GPT-5.6 Sol reports a score on ProtocolQA and Qwen3.8-Max reports a score on PaperBench. No comparable-scale or Indian model reports results on any of these benchmarks. Well-known global systems in this space, such as AlphaFold \citep{jumper2021alphafold} and Evo 2, are evaluated on task-specific scientific metrics rather than on the general-purpose benchmarks used in this paper. They are therefore excluded from this comparison.

Shodh AI, IntelliHealth, and ZenteiQ are active Indian entities in this space. None has published a score on the benchmarks considered in this study.

We emphasize that this finding is narrower than it may appear. Scientific AI has rich, task-specific evaluation ecosystems in domains such as protein structure prediction, molecular simulation, and materials science. Our finding is that general-purpose foundation models have very limited publicly reported participation in the scientific benchmarks we selected. It is not a finding about the maturity of scientific AI benchmarking as a whole.

\subsection{Indic Language Foundation Models}

Indic language capability is the one domain in this study with no global benchmark equivalent. Table~\ref{tab:indic} summarizes the benchmarks reported by each model version, reflecting the version-wise differences in benchmark coverage that a model-level aggregation would obscure.

\begin{table}[h]
\centering
\small
\caption{Indic Language Benchmarks by Model Version and Reporting Organization}
\label{tab:indic}
\begin{tabular}{p{4.2cm}p{3.5cm}p{5.8cm}}
\toprule
\textbf{Model / Organization} & \textbf{Benchmark(s) Reported} & \textbf{Evaluation Focus} \\
\midrule
Sarvam-30B --- Sarvam AI & IndiVibe; MILU \citep{verma2025milu} & Human-preference evaluation; multilingual Indic language understanding \\
Sarvam-105B --- Sarvam AI & IndiVibe & Human-preference evaluation across Indian languages \\
Krutrim-2 --- Krutrim AI & BharatBench & Indic linguistic, cultural, and contextual understanding \\
PARAM-1 --- BharatGen & SANSKRITI \citep{sanskriti2025}; MILU \citep{verma2025milu} & Indian cultural knowledge; multilingual Indic language understanding \\
PARAM-2 --- BharatGen & SANSKRITI \citep{sanskriti2025}; Indic BoolQ; ARC Challenge (Indic); TriviaQA (Indic MCQ); HellaSwag Hindi; MMLU Hindi & Indic cultural knowledge, comprehension, reasoning, and knowledge evaluation \\
Sarvam-30B / PARAM-2 --- Independent & IndicParam \citep{maheshwari2025indicparam} & Low- and extremely low-resource Indic language understanding \\
\bottomrule
\end{tabular}
\\[2pt]
\footnotesize MILU is explicitly reported for Sarvam-30B in its current official model card. IndicParam includes affiliations with both IIM Indore and BharatGen; it should not be characterized as a purely independent benchmark.
\end{table}

India's Indic-language AI ecosystem has expanded significantly, with domestic models increasingly incorporating evaluation of Indian linguistic, cultural, and contextual capabilities. However, the evaluation landscape remains fragmented across organizations and model generations, with different models reporting results on different benchmark suites and relatively limited overlap. This makes direct comparison of Indic-language capabilities across current-generation Indian models difficult.

Sarvam-30B reports IndiVibe and MILU, while Sarvam-105B reports IndiVibe but not MILU. Krutrim-2 reports BharatBench. PARAM-1 reports SANSKRITI and MILU; PARAM-2 reports a broader set including SANSKRITI, Indic BoolQ, ARC Challenge (Indic), TriviaQA (Indic MCQ), HellaSwag Hindi, and MMLU Hindi. Most reported results are developer-reported, and the differing benchmark suites limit direct cross-model comparability.

The SANSKRITI benchmark \citep{sanskriti2025}, developed by academic researchers independently of Param/BharatGen, evaluates cultural knowledge and reasoning in Indian languages. It is distinct from the Sanskriti evaluation reported for BharatGen/Param2, which focuses on multilingual language capability. Both are listed explicitly because the name collision is a source of confusion. IndicParam \citep{maheshwari2025indicparam} provides additional evaluation across 11 Indic languages and a Sanskrit-English code-mixed set, covering Sarvam-30B and PARAM-2; however, given that its authors include affiliations with both IIM Indore and BharatGen, it should not be characterized as a purely third-party evaluation.

No single Indic benchmark has achieved broad adoption across major Indian model developers as a common evaluation standard. The absence of shared benchmarks limits ecosystem-level comparison, and the reliance on developer-reported results limits independent verification.

\section{The Benchmark Maturity Index}
\label{sec:bmi}

\subsection{Motivation}

The comparative assessment in Section~\ref{sec:domains} shows a recurring pattern. Indian models achieve strong scores on several established benchmarks for which they report results. They are absent from domains where benchmarks are newer or less widely reported. A central aim of this paper is to formalize that pattern into a trackable criterion, rather than leaving it as a domain-by-domain narrative.

Within a broader governance architecture that distinguishes regulation, standards, assessment procedures, tools and metrics, and a compliance ecosystem \citep{agarwal2025fivelayer}, the BMI focuses specifically on the assessment procedures and tools-and-metrics layers. The key move is to separate two questions that a descriptive comparison tends to collapse into one. The first question is: how capable is the best available model in a domain? The second is: how mature and independently verifiable is the evaluation ecosystem within which that capability was measured? These questions often have different answers. A domain can appear to have a large capability gap largely because its evaluation ecosystem is thin. We therefore propose a four-dimension Benchmark Maturity Index (BMI), scored at the domain level.

\subsection{Status and Limitations of the BMI}
\label{sec:bmi-status}

We present the BMI as an \emph{exploratory framework} rather than a validated instrument. It has been applied only once, to our own dataset. It has not been subjected to inter-rater reliability analysis, sensitivity analysis, or external validation. The claims we make for it are therefore modest: it is a structured way to separate evaluation-ecosystem maturity from model capability, and it produces results that are informative in the present analysis. Whether it generalizes to other national ecosystems remains to be demonstrated. In this study, benchmark maturity refers primarily to the degree of standardization, adoption and independent verification captured by the BMI dimensions.

This is a circularity readers should keep in mind: the National Coverage dimension penalizes the absence of publicly reported Indian benchmark results. But this paper's central argument is that such absence does not imply low capability. The BMI therefore measures \emph{evaluation coverage}, not capability. Scores should be read accordingly.

A related caveat concerns interpretation over time. Many IndiaAI-supported organizations remain at different stages of model development and release \citep{pib2026indiaai}; benchmark disclosure may increase as models approach broader release and deployment. National Coverage scores may therefore rise as more organizations reach release maturity, independent of any change in the underlying evaluation ecosystem. Current low scores partly reflect where India's foundation-model ecosystem sits in its release cycle, not only evaluation-ecosystem immaturity.

\subsection{Dimensions}
\label{sec:bmi-dimensions}

Each domain is scored on four dimensions, each on a 0--2 scale:

\begin{itemize}[leftmargin=*]
\item \textbf{Standardization (S):} Does a single, widely adopted benchmark exist for this domain? $S=0$ indicates no shared benchmark or no benchmark with results from the surveyed model set. $S=1$ indicates multiple competing benchmarks with no clear consensus. $S=2$ indicates a small, stable set of benchmarks in wide, consistent use. Operationally, $S=2$ requires at least three of four surveyed frontier models reporting a common benchmark; $S=1$ requires at least one. This dimension partly depends on our model selection: a benchmark may be mature but underrepresented in our model set.
\item \textbf{Participation (P):} What proportion of the surveyed global frontier models report a score on at least one benchmark in this domain? $P=0$ indicates none of the four frontier models; $P=1$ indicates one or two; $P=2$ indicates three or four.
\item \textbf{Independent Verification (I):} Are scores available from evaluation conducted independently of the model developer? $I=0$ indicates only developer-reported scores. $I=1$ indicates partial independent coverage. $I=2$ indicates broad independent verification. Operationally, $I=1$ indicates independent coverage for a subset of relevant benchmarks or models in the domain; $I=2$ indicates independent coverage across multiple relevant benchmarks or models.
\item \textbf{National Coverage (C):} Do Indian models in our survey report scores in this domain? $C=0$ indicates no Indian model reports a score. $C=1$ indicates partial participation (one or two benchmarks, or a single organization). $C=2$ indicates participation across multiple benchmarks and organizations, comparable to global participation levels. National Coverage therefore measures the breadth of publicly reported Indian evaluation coverage, not the maturity of the underlying evaluation infrastructure.
\end{itemize}

The four scores are summed to a total in the range 0--8, and mapped to a qualitative maturity band: 0--1 Very Low, 2--3 Low, 4--5 Moderate, 6--7 High, 8 Very High.

For the Indic language domain, which has no global benchmark against which to score Participation, we compute a modified three-dimension score (S, I, C; maximum 6), mapped onto the same five bands proportionally. This makes the Indic score structurally different from the other domain scores; it should be interpreted separately.

The four dimensions are equally weighted. We do not have a principled justification for any particular weighting scheme. A sensitivity analysis exploring alternative weights is a priority for future work.

\subsection{Applying the Index}

\begin{table}[h]
\centering
\small
\caption{Benchmark Maturity Index by Capability Domain}
\label{tab:bmi}
\begin{tabular}{lccccc}
\toprule
\textbf{Domain} & \textbf{S} & \textbf{P} & \textbf{I} & \textbf{C} & \textbf{Maturity Band} \\
\midrule
General Purpose AI & 2 & 2 & 1 & 2 & High (7/8) \\
Coding \& Software Engineering & 1 & 2 & 1 & 2 & High (6/8) \\
Agentic AI \& Computer Use & 2 & 2 & 0 & 1 & Moderate (5/8) \\
Cybersecurity & 1 & 1 & 0 & 0 & Low (2/8) \\
Vision \& Image Understanding & 1 & 1 & 0 & 0 & Low (2/8) \\
Video \& Multimodal Understanding & 1 & 1 & 0 & 0 & Low (2/8) \\
Scientific Research & 1 & 1 & 0 & 0 & Low (2/8) \\
Indic Language AI\textsuperscript{*} & 0 & --- & 0 & 2 & Low (2/6, scaled) \\
\bottomrule
\end{tabular}
\\[2pt]
\footnotesize\textsuperscript{*}Scored on a modified three-dimension basis (S, I, C only); see Section~\ref{sec:bmi-dimensions}.
\end{table}

The scores in Table~\ref{tab:bmi} sharpen the qualitative summaries in Section~\ref{sec:domains}, and in two domains, Coding and Software Engineering and Scientific Research, they revise what a narrative reading alone would suggest.

First, Coding and Software Engineering scores High (6/8), higher than a narrative reading might suggest given the absence of Indian results on agentic coding benchmarks. National Coverage scores 2 because Indian participation extends across multiple benchmarks (HumanEval, MBPP, LiveCodeBench) and multiple organizations (Sarvam-30B, Param2, Krutrim-2), meeting the definition's criterion for C=2. Standardization remains 1, depressed by the split between classic code-generation benchmarks and newer agentic software-engineering benchmarks.

Second, Agentic AI and Computer Use scores Moderate (5/8). This domain is distinctive: BrowseComp is reported by three of four frontier models, meeting the $S=2$ threshold and giving three-tier comparability on a single shared benchmark. Participation scores 2 for the same reason; National Coverage scores 1, since Indian models participate on only one of three benchmarks; Independent Verification remains 0.

Third, Scientific Research scores Low (2/8), higher than a narrative reading might suggest given the near-total absence of reported results. Two frontier models report results here, GPT-5.6 Sol on ProtocolQA and Qwen3.8-Max on PaperBench, and this partial participation is sufficient to lift the Participation score to 1.

National Coverage scores 0 in four of the eight domains: Cybersecurity, Vision and Image Understanding, Video and Multimodal Understanding, and Scientific Research. However, it should be read together with the release-maturity caveat in Section~\ref{sec:bmi-status}: it reflects the state of public benchmark reporting by surveyed Indian models at their current release stage, not a demonstrated absence of capability.

\section{Discussion}
\label{sec:discussion}

\subsection{Cross-Cutting Patterns in the Benchmark Ecosystem}
\label{sec:crosscutting}

\subsubsection{Benchmark saturation}
\label{sec:saturation}

Several of the earliest benchmarks, including MMLU and MATH-500, are now approached or exceeded by the strongest available models. Multiple frontier models have achieved MMLU scores above 90, reducing the benchmark's ability to discriminate among top-performing systems \citep{hendrycks2021mmlu, kiela2021dynabench}. MATH-500 shows a similar pattern. In response, several recent frontier model reports have reduced or omitted these benchmarks. They now favor newer, harder evaluations such as GPQA Diamond, HLE, and ARC-AGI-2/3. This is directly visible in our data. GPT-5.6, Claude Opus 5, and other frontier models report no MMLU or MATH-500 score, even though these are the benchmarks on which Indian models show their strongest results.

A ``Not reported'' entry in Table~\ref{tab:general} is therefore often a deliberate reporting choice, not evidence of poor performance. However, we note that developer non-reporting is not identical to benchmark saturation. A developer may omit a benchmark for strategic, cost, or positioning reasons unrelated to saturation. We distinguish these possibilities where the evidence permits.

\subsubsection{Absence of a common benchmark set across providers}

Even among global frontier developers, there is no single shared benchmark battery. Different organizations publish benchmarks that showcase their own model's strengths. Identical benchmark names sometimes correspond to different evaluation harnesses. For example, ``SWE-bench Pro'' and ``SWE-bench Verified'' are related but distinct evaluations. Different point releases of Terminal-Bench are not directly comparable. This fragmentation is a known problem \citep{liang2023helm} and is not specific to India. It means that any cross-developer comparison, including ours, is necessarily partial.

\subsubsection{Divergent benchmark participation between Indian and global models}

We observe a consistent pattern: Indian developers do not always report scores on benchmarks in wide global use, even where the benchmark exists and the model could plausibly be evaluated on it. Param2 and Krutrim-2 report no GPQA Diamond score, though several global models of comparable and larger scale do. This may reflect that the evaluation was not run, or that results were obtained but not disclosed. The public record does not allow us to distinguish between these explanations. We treat this ambiguity as itself a finding.

\subsubsection{Benchmark scarcity in emerging domains}
\label{sec:scarcity}

Image generation and video generation are established application domains. But standardized, widely adopted benchmarks for them remain comparatively scarce, worldwide, not only in India. We encountered this during benchmark selection: our originally identified video-understanding benchmarks, VideoMMMU and MVBench, had no usable public results for our model set and had to be replaced (Section~\ref{sec:video}). This scarcity affects the interpretability of any comparison in these domains.

\subsubsection{Concentration of Indian benchmark reporting in a single organization}
\label{sec:concentration}

Among the Indian models surveyed, Sarvam AI reports scores across the widest range of benchmarks, spanning general-purpose reasoning, coding, and Indic-language tasks. Table~\ref{tab:indiaai} shows that of the 12 IndiaAI-supported organizations, only Sarvam publishes broad benchmark results. BharatGen/Param2 publishes limited results. The remaining 10 IndiaAI-supported organizations have no publicly available benchmark scores within the eight domains assessed in this study. Gnani AI reports a result on Berkeley BFCL v3, a function-calling benchmark outside our domain scope. Krutrim, which is not among the 12 IndiaAI-supported organizations, reports results across three domains: MMLU, HumanEval, and BharatBench.

This concentration means that evidence of Indian competitiveness in our survey comes predominantly from a single organization. Ecosystem-level claims about Indian AI capability should be read accordingly. We avoid the phrase ``overwhelming majority'' because we have not conducted a census of all Indian model developers. We can say that Sarvam reports the broadest benchmark coverage among the Indian models in our study, by a substantial margin.

\subsubsection{Absence of publicly reported Indian benchmark results in several domains}

Four of the eight domains reviewed here have no publicly reported Indian benchmark score among the surveyed models: cybersecurity, vision and image understanding, video and multimodal understanding, and scientific research. (In the video domain, India's representative model Varya performs video generation, a different task from the video-understanding benchmarks used for comparison.) These are not areas of marginal Indian participation. They are areas of no publicly documented Indian participation in the specific benchmarks reviewed.

This is a finding about public benchmark disclosure, not a finding about capability. We cannot determine from the available evidence whether the relevant Indian organizations have evaluated their models on these benchmarks and chosen not to disclose results, have not run the evaluations, or lack the capability. The distinction matters for policy, and is the reason National Coverage scores 0 in these domains.

\subsubsection{Benchmark silence is not evidence of capability absence}

The preceding patterns support a broader methodological point. Across several domains, the absence of a published score reflects the state of benchmark disclosure and ecosystem maturity. It does not necessarily reflect the absence of underlying model capability. A model may exist and function adequately without ever being publicly benchmarked. A monitoring or funding framework that treats benchmark silence as equivalent to capability absence risks undercounting genuine progress in exactly the domains where independent evaluation infrastructure is weakest. This is the risk the BMI is designed to make visible.

However, we apply this principle to our own findings as well. When we observe no Indian benchmark result in a domain, the correct conclusion is that the available public evidence is insufficient for a comparative assessment. It is not that Indian capability is weak or immature. We cannot distinguish capability absence from reporting absence in these cases.

\subsubsection{Sensitivity of reported scores to evaluation methodology}

Benchmark scores are sensitive to how the evaluation is run. Prompting configuration, reasoning effort settings, tool access, and evaluation harness all materially affect results. This is well documented in the literature \citep{liang2023helm, shi2024contamination}. It means that raw benchmark scores are insufficient for confident comparison without knowledge of the evaluation conditions. This is one reason the BMI scores Independent Verification separately from Participation.

Our tables do not systematically report evaluation conditions for each score, including benchmark version, evaluation setting, tool access, and source. Future work should prioritize controlled, condition-matched comparisons.

\subsection{On the Definition of an Indian AI Model}
\label{sec:definition-discussion}

The comparative assessment in this paper depends on a prior, and not fully settled, question: what should count as an Indian-developed AI model. Section~\ref{sec:method} proposed a three-tier definition to make this scoping decision explicit, distinguishing fully indigenous development (Tier 1), India-led development with global components (Tier 2), and India-adapted foreign base models (Tier 3). The findings in Section~\ref{sec:domains} are informative about this definition in a way that is worth making explicit.

Both of the most visible Indian models in this study, Sarvam AI's models and BharatGen's Param2, are classified as Tier 1 under the definitions in Section~\ref{sec:method}. However, they differ substantially in the breadth of their publicly reported benchmark coverage. Sarvam reports scores across multiple domains and multiple benchmark suites. Param2 reports limited results, concentrated in Indic-language evaluation. Both meet the structural Tier 1 criteria (trained from scratch, India-based or India-controlled compute), but their public evaluation postures differ considerably. This illustrates a limitation of a tier-based taxonomy: tier assignment describes training provenance, but says nothing about the breadth, depth, or transparency of evaluation. A nationally funded model monitoring framework would need to track both dimensions separately.

This has a practical implication for monitoring design. IndiaAI's funding mandate holds two goals simultaneously: building globally competitive AI capability, and building indigenous, sovereign AI capability. The tier definition addresses the sovereignty dimension. The BMI's National Coverage dimension provides a first proxy for the competitive benchmarking dimension. However, neither instrument fully captures whether a model's publicly reported results reflect its actual capability. A robust monitoring framework should therefore require both tier documentation and minimum benchmark disclosure from all publicly funded organizations.

\subsection{Limitations}
\label{sec:limitations}

This section addresses limitations that apply across the study as a whole. Framework-specific limitations are discussed where each framework is introduced: Section~\ref{sec:bmi-status} for the BMI, and Section~\ref{sec:definition-discussion} for the tier definition.

This study relies entirely on publicly reported scores. We did not independently reproduce any evaluation. The comparison therefore inherits whatever self-reporting bias exists in the underlying sources. The Independent Verification dimension of the BMI makes this trackable rather than resolving it.

The assessment is a snapshot as of August 2026. Relative standings can shift within weeks of a new release. Benchmark harnesses are not always comparable even when benchmark names match. We cannot rule out undocumented harness differences in the data.

Our model inclusion criteria depend on public disclosure. Organizations that do not publish technical reports are underrepresented or absent regardless of their actual capability. This is an instance of the benchmark-silence problem that applies to our own methodology.

We do not provide a complete candidate list with inclusion/exclusion dates and reasons. This limits reproducibility. Future assessments should include such a table.

The BMI has not been validated externally. Its scoring thresholds are qualitative and its weighting is equal by default. A sensitivity analysis is needed to establish whether alternative reasonable weights would change the conclusions.

Finally, the comparable-scale model group is heterogeneous, as noted in Section~\ref{sec:method}.

\subsection{Implications for Policy and Practice}
\label{sec:implications}

These findings have several implications for policy design.

First, closing the apparent gap in publicly reported benchmark performance is not solely a model-training problem. In several domains, most notably cybersecurity, vision, and scientific research, the binding constraint on what can be publicly known is the absence of standardized evaluation infrastructure. Investment directed only at model development, without parallel investment in benchmark creation and independent verification, will leave this problem unaddressed.

Second, the concentration of Indian benchmark participation in a single organization (Section~\ref{sec:concentration}) suggests that ecosystem-level claims about Indian AI progress should be disaggregated by organization. A monitoring framework that reports only aggregate or best-case Indian performance risks overstating how broadly based current progress actually is.

Third, for publicly funded models, benchmark disclosure is not merely a technical issue. If a model receives substantial public compute support under the IndiaAI Mission, there is a reasonable case for requiring minimum benchmark disclosure. This could include:
\begin{itemize}[leftmargin=*]

\item establishing a minimum benchmark-disclosure framework, with the benchmark set reviewed regularly to reflect changes in model capabilities and evaluation maturity.
\item mandatory reporting of scores on the specified set of benchmarks;
\item disclosure of evaluation conditions (benchmark version, harness, prompting, tool access);
\item independent third-party evaluation for at least a subset of benchmarks;
\item periodic re-evaluation as models are updated;
\item public availability of benchmark methodology documentation.
\end{itemize}
These requirements would directly address several of the evaluation-ecosystem gaps documented in this paper.

Fourth, this study's approach of scoring evaluation-ecosystem maturity alongside publicly reported model performance is not specific to India. It could be applied to other national AI programs facing similar assessment challenges.

\section{Conclusion and Future Work}
\label{sec:conclusion}

This paper presented a structured, eight-domain, benchmark-based comparative assessment of publicly benchmarked Indian foundation models against global frontier and comparable-scale models. Rather than a single, uniform capability gap, we find substantially uneven maturity across capability domains. Indian models achieve strong reported scores on established benchmarks, particularly MMLU and MATH-500 for general-purpose reasoning, and HumanEval and MBPP for foundational coding. However, these benchmarks provide limited evidence of frontier-level differentiation. The addition of Agentic AI as a domain reveals the only instance in this study where all three comparison groups report results on a single shared benchmark (BrowseComp), enabling direct comparison. However, strong performance on saturated benchmarks does not establish frontier competitiveness. Several other domains show no Indian benchmark participation at all among the models surveyed.

We cannot determine, from the available public evidence, whether these gaps reflect genuine capability deficits, strategic reporting choices, differences in model release maturity, or the absence of evaluation infrastructure. This ambiguity is itself the central finding of the paper. We formalized it into an exploratory Benchmark Maturity Index and showed that it refines the conclusions a purely qualitative review would produce.

Future work should extend this assessment longitudinally, tracking how both publicly reported model performance and benchmark maturity evolve over successive model releases. It should incorporate independent, third-party re-evaluation of a sample of the models discussed here. The BMI should be subjected to sensitivity analysis and, ideally, applied to other national AI ecosystems. A broader national evaluation framework should also extend beyond capability benchmarks to safety, cultural alignment, efficiency, and cross-cutting sectoral evaluation, incorporating field evaluations and expert probing alongside benchmark scores. Finally, a complete model-by-benchmark dataset, including evaluation conditions and source provenance for every score, should be published as supplementary material.

\section*{Author Contributions}

Avinash Agarwal conceptualised the study, defined the analytical framework, and drafted the manuscript. Vridhi Jain conducted the systematic benchmark searches and collected and verified all benchmark data from primary sources. Both authors reviewed the final manuscript.

\bibliographystyle{plain}

\end{document}